# Interplay of symmetry-conserved tunneling, interfacial oxidation and perpendicular magnetic anisotropy in CoFeB/MgO-based junctions

Pravin Khanal, Bowei Zhou, Hamid Almasi, Ali Habiboglu, Magda Andrade, Jack O'Brien, Arthur Enriquez, Carter Eckel, Christopher Mastrangelo and Wei-Gang Wang*

Department of Physics, University of Arizona, Tucson, AZ 85721, USA

The interfacial oxidation level and thermodynamic properties of the MgO-based perpendicular magnetic tunneling junctions are investigated. The symmetry-conserved tunneling effect depends sensitively on the MgO adatom energy during the RF sputtering, as well as the thermal stability of the structure during the post-growth thermal annealing. Two different failure modes of the magnetoresistance are highlighted, involving with the decay of perpendicular magnetic anisotropy and destruction of coherent tunneling channels, respectively. Through the careful control of interfacial oxidation level and proper selection of the heavy metal layers, both perpendicular magnetic anisotropy and tunneling magnetoresistance of the junctions can be increased.

*wgwang@arizona.edu

Spin-dependent tunneling[1] has been under intensive investigation in the past three decades due to its intriguing physics and the wide application of magnetic tunnel junctions (MTJs). [2,3] The core structure of an MTJ consists of an very thin insulator ( ~1nm) sandwiched by two ferromagnetic (FM) thin films. The transmission probability of electron wavefunctions depends on the relative magnetization orientations of the two FM layers in the structure. The tunneling magnetoresistance (TMR) characterizes the conductance ratio of an MTJ in the parallel (P) and antiparallel (AP) configurations, and is generally proportional to the spin-polarization (SP) of the FM layers.[2] Modern MTJs employ the so-called symmetry filtering tunneling ( or symmetry-conserved tunneling) effect to generate large TMR ratios.[4] Generally speaking, the Bloch electrons in *3d* transitional metals can be classified by their symmetries into different bands. Take Fe for example, though the SP due to the total density of states at the Fermi energy is only about 50%, the SP of the $\Delta_1$ band electrons is 100%. This feature, combined with materials such as MgO that offers a much higher transmission rate to $\Delta_1$ band electrons compared to electrons with other symmetries (such as $\Delta_5$ and $\Delta_{2'}$), gives rise to the symmetry-filtering effect[4] where giant TMR up to 600% can be obtained at room temperature (RT).[5] The maximal TMR with amorphous AlOx barrier is around 80%,[6] so generally such symmetry filtering effect exists in MTJs with TMR larger than 100%.

Of particular interest are the MTJs with perpendicular magnetic easy axis (pMTJs). [7–9] The perpendicular magnetic anisotropy (PMA) in these junctions can originate from the FM/oxide interfaces, therefore it can be modulated by the electric field applied to the tunnel barrier through the voltage-controlled magnetic anisotropy (VCMA) effect. [10,11] VCMA offers a unique approach to achieve ultralow energy switching of MTJs compared to spin-transfer torques (STT)[12,13] and spin orbit torques (SOT)[14] based switching scenarios. Indeed, low energy switching of a few fJs has been demonstrated in pMTJ by the VCMA effect, [15–18] which is at least one order of magnitude lower than that of the STT or SOT.

Due to the much thinner thickness of the FM layers used these junctions, typically ~1nm of CoFeB, [7,19,20] the TMR of pMTJs are substantially smaller than that of those with in-plane magnetic anisotropy. Here we investigate the influence of oxidation level at the CoFeB/MgO interface and PMA of the heavy-metal(HM)/CoFeB/MgO structure on the tunneling magnetoresistance of the pMTJs. The behaviors of pMTJs after high temperature annealing were investigated in different junctions. Through controlling the adatom energy during the deposition of MgO barrier and optimization the PMA of the junctions, an enhanced TMR of 257% was obtained in the voltage controllable pMTJs.

The MTJ films in this work were fabricated in a 12-source UHV sputtering system (AJA International) with a base pressure of $10^{-7}$ Pa ($10^{-9}$ Torr). The stack structure of the films is Si/SiO$_2$/Ta(8 nm)/Ru(7 nm)/ HM$_{BOT}$/Co$_{20}$Fe$_{60}$B$_{20}$(1nm)/MgO(1.2-3.5nm)/ Co$_{20}$Fe$_{60}$B$_{20}$(0.9-1.5nm)/ HM$_{TOP}$/Ru(15nm), where HM$_{BOT}$ and HM$_{TOP}$ are the heavy metals layer with the Ta(5nm)/Mo (0.7nm) and Mo (1.2nm)/Ta (10nm), respectively, unless otherwise noted. The schematic structure of the films is shown in Fig. 1(a). Circular junctions with diameters ranging from 2μm nm to 100 μm were patterned and subsequently annealed under varying conditions. Detailed information on sample fabrication and characterization can be found in our previous publications.[21–24]

The large TMR in the CoFeB(001)/MgO(001)/CoFeB(001) MTJs rely on the pristine CoFe(001)/MgO(001) interface after Boron atoms diffusing out the originally amorphous CoFeB layers during annealing. [25,26] It has been revealed by DFT calculations that if the Fe/MgO interface is contaminated, for example, by excessive amount of Oxygen atoms, the TMR would reduce dramatically. [27] This is largely due to the reduced transmission of the $\Delta_1$ band electrons in the majority spin channel of the parallel configuration, even the transmission of the majority-minority channels involving $\Delta_5$ and $\Delta_{2'}$ bands stay unchanged. In fact, the TMR can be reduced by more than 10 times in an over-oxidized junction when one monolayer of Oxygen was placed between Fe and the MgO layers in the DFT calculation. [27] Experimentally, though it is well expected that over-oxidation of Fe or CoFe is detrimental to TMR, it is unclear such FeO layers still exist in MTJs with hundreds of percent of TMR. The exchange-bias effect caused by such an interfacial FeO layer was previously measured by the magnetization-induced second harmonic generation (MSHG) technique[28], however no magnetoresistance was investigated in this study so it was still not clear that such FeO layer can be a reason to destruct the transmission of $\Delta_1$ electrons in a high quality MTJs. In a more recent study by us, the existence of such FeO layer was revealed by the shift of the TMR curves measured at low temperature after field cooling, in high quality pMTJs with 400% TMR at 30K.[21] Furthermore, the excessive amount of O at the CoFeB/MgO interface was also demonstrated by EELS measurement. Through the fitting to Malozemoff model[29] the antiferromagnetic anisotropy constant was determined to be $10^4$ J/m$^3$,, which is about 1% of the typical value found in the FM/FeO bilayers.[30]. These results unambiguously demonstrate an interfacial FeO oxide layer can exist in modern pMTJs with high TMR.

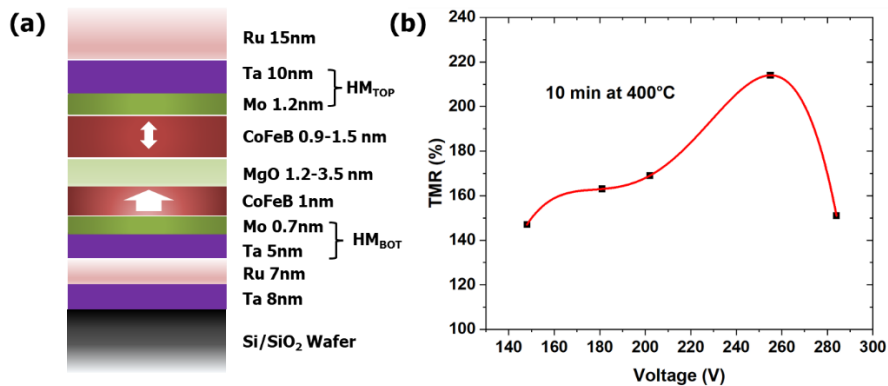

Figure 1. (a) Schematic structure of the MTJs explored in this study. (b) TMR of a series of pMTJs where the MgO barriers were fabricated under different RF sputtering voltages.

This FeO layer, even likely only exist as discontinuous patches at the CoFe/MgO interface instead of a continuous layer, is expected to impede the transmission of $\Delta_1$ electrons thus reducing the symmetry filtering effect in the junctions. Therefore it is highly desirable to eliminate or decrease this negative impact. Compared to other thin film fabrication methods, magnetron sputtering is known to be a high energy process. It is not a surprise that the bottom CoFeB could get oxidized upon the impinging of the high-energy MgO flux. Generally speaking, this interfacial FeO layer can be potentially controlled by a number of methods, such as tunning the MgO adatom energy during deposition, inserting an ultrathin "dusting" Mg layer prior the deposition of MgO, [31,32] varying Ar pressure during MgO deposition,[6,33] and reducing FeO to Fe in the post-growth thermal annealing.[34] To investigate the effect of the MgO adatom energy during deposition, a series of samples has been fabricated, in which MgO deposition voltage was varied from 150V to 290V. The MTJs were all annealed at 400°C for 10min prior testing. The behavior of the TMR is plotted in Figure 1(b). A couple of interesting features immediately emerge. First the TMR ratios tends to be smaller with lower deposition voltage. This is in agreement with the common understanding that a high level of impurities and/or defects [35,36] often develops in films fabricated by sputtering if the deposition rate is not high enough. Another reason can be that the crystalline quality of MgO deposited at lower energy is not as good, since the templating effect of the (001)-oriented MgO in the as-deposited state is critical to the solid state epitaxy of the CoFe(001)/MgO(001)/CoFe(001) heterostructure[25] that affords the symmetry filtering effect. This is why in most cases high voltage around 300V is used during the RF deposition of MgO. Second, the TMR starts decreasing when the deposition voltage is at 285V, suggesting the formation of FeO becomes more pronounced under this condition as the kinetic energy of $Mg^{2+}$ and $O^{2-}$ adatoms is getting higher. The occurrence of maximum TMR at 255V indicates that a good balance of the MgO quality and FeO formation is achieved under this condition. Lastly it should be noted that the best voltage (255V) corresponding to the highest TMR in Figure 1(b) only holds for annealing carried out at 400°C or above. Indeed the TMR is higher in the samples deposited 200V if the annealing is carried out at 340°C ( not shown). This fact indicates that tuning of interfacial oxidation level by varying MgO deposition voltage needs to be done with the consideration of desirable annealing conditions, which is in agreement with the early study showing that interfacial FeO in as-deposited samples generally gets partially reduced during the thermal annealing process.[34]

These results demonstrate that the FeO layer can be effectively reduced by tunning the adatom energy of MgO during RF sputtering. Another method to reduce the interfacial oxidation is adding a very thin Mg layer prior the deposition of MgO.[31,32] The lowest thickness of Mg film that can be reproducibly achieved by our deposition tool is about 1Å. Therefore 1Å of Mg was inserted between the CoFeB layer and MgO layer to investigate if further increase of TMR can be achieved. Two representative TMR curves for the pMTJs without and with 1Å Mg insertion are shown in Figure 2. The MgO layer of these two types of pMTJ was both deposited at 255V and the pMTJs were annealed under the identical condition ( 30min at 400°C ). It can be seen that the incorporation of 1Å Mg didn't help to increase the TMR. On the contrary, both TMR and PMA are smaller compared to the pMTJ without the Mg insertion. The PMA at the CoFeB/MgO interface originates for the hybridization of 3d-2p wavefunctions of Fe (Co) and O. It is known from DFT calculations that an over-oxidized or under-oxidized interface is not able to produce a large PMA as the optimally oxidized interface.[37] The TMR has a very similar dependence on over/under-oxidation level of the CoFe/MgO interface, as demonstrated by the enhanced TMR by reducing FeO in Fig. 1(b). To further exclude the effect of PMA in understanding the impact of Mg insertion to symmetry filtering tunneling, MTJ with in-plane magnetic anisotropy were fabricated and tested. The core structure of these samples is CoFeB (6nm)/MgO (2.5nm)/CoFeB (4nm) . As shown in Figure 2(c), there is no obvious difference between the two types of samples. Therefore it is concluded that with the already optimized MgO deposition voltage of 255V, the Mg insertion cannot further enhance the coherent tunneling in the MTJs.

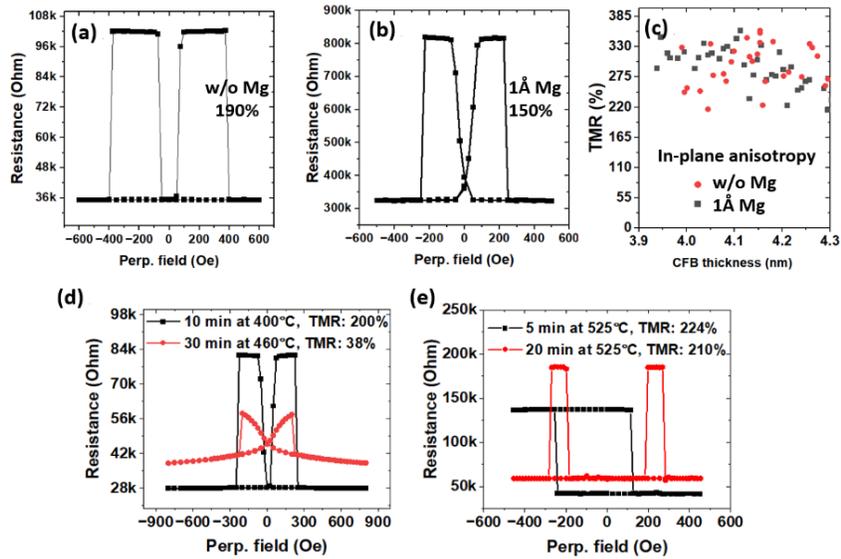

Figure 2. (a) Representative TMR curve of the pMTJ without Mg insertion. (b) Representative TMR curve of the pMTJ with 1Å Mg insertion between the bottom CoFeB electrode and MgO barrier. (c) Comparison of TMR in in-plane MTJs with and without Mg insertion. (d) TMR curves of pMTJ with weak PMA after two successive annealings. e) TMR curves of pMTJ with strong PMA after two successive annealings.

One of the most essential steps to achieve high TMR in the CoFeB/MgO/CoFeB MTJ is the post-growth thermal annealing.[25] In addition to tunning the interfacial oxidation level as discussed, the symmetry-conserved tunneling effect is established during the annealing process, as reflected the development of the $\Delta_1$, $\Delta_{2'}$ and $\Delta_5$ bands through the crystallization of the amorphous CoFeB.[26,38,39] It is the in this process the TMR grow from tens of percent to hundreds of percent. The dynamic evolution of the symmetry filtering tunneling during annealing of MTJs with in-plane anisotropy involves a three-phase process, where TMR generally increases with the rapid crystallization of CoFeB ( phase-1) and crystallinity improvement of MgO (phase-2), followed by the subsequent reduction (phase-3) due to the destruction of $\Delta_1$ channel as reflected by the decrease of the P state conductance.[26] Therefore it is essential to understand under what annealing condition maximum TMR can be obtained in pMTJs. Two different mechanisms are identified that could lead to the reduction of TMR during annealing. The first one is the loss of AP state due to degradation of PMA.[40] One such example is shown in Figure 2 (e), where an MTJ was successively annealed under different conditions. It can be clearly seen that the AP plateau of the TMR curve after 30min of annealing at 460°C no longer exists, therefore the TMR ratio is much reduced. The second mechanism for TMR reduction in pMTJ is of the transport nature, due to the destruction of the coherent tunneling channels. As Shown in Fig. 2(e), the AP plateau in the MTJ after 20min annealing at 525°C is well maintained, however the TMR is still reduced. This decrease of TMR is related to that the reduction of majority-minority transmission involving the $\Delta_5$ and $\Delta_{2'}$ electrons (reflected by increase of AP resistance) is outpaced by the reduced transmission of $\Delta_1$ electrons (reflected by increase of P resistance) due to the increased defect scattering, similar to what was observed in the MTJs with in-plane anisotropy.[26]

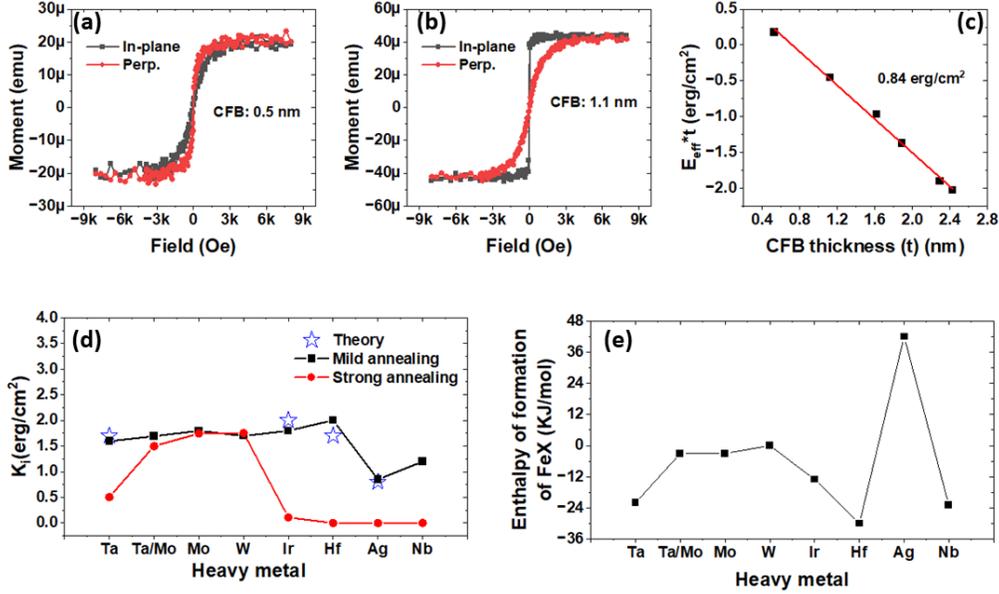

Figure 3. (a) Hysteresis loops of the Ag (3nm)/CoFeB (0.5nm)/MgO (2.5nm). (b) Hysteresis loops of the Ag (3nm)/CoFeB (1.1nm)/MgO (2.5nm). (c). Determination of $K_i$ value for the Ag/CoFeB/MgO structures. (d) Comparison of $K_i$ values with HM/CoFeB/MgO structures where the HM are Ta, Mo,[41] Ta/Mo,[20] W,[19] Ir, Hf, Ag and Nb.[42] (e). Calculated values of enthalpy of formation for different HM-Fe alloys, adapted from Ref. 43.

From these results it is clear that a strong PMA is critical to high TMR. Ideally, the limitation of TMR should be the destruction of coherent tunneling channels, as shown in Fig. 2(e), instead of decay of PMA in Fig. 2(d). The PMA of the HM/CFB/MgO structure depends on few factors such as the oxidation level at the CoFeB/MgO interface, the formation energy of the HM element with Fe and Co, and the formation energy of HM with B etc.[44] Particularly, it was identified that the formation energy of between HM and Fe plays an important role.[20] As a result, Mo and W gives rise to relatively high PMA after high temperature annealing.[19,20] Another element, Ag, was predicted to have very high formation energy with Fe,[43] but no pMTJ with Ag as HM has been reported previously. A series of films with the structure of wafer/Ag (3nm)/CoFeB (0.5-2.5nm)/MgO (2.5nm) were fabricated and the dependence of magnetic anisotropy on the thickness of CoFeB was investigated by VSM. Indeed the samples exhibit perpendicular easy axis when then CoFeB thickness is small, as shown in Fig. 3(a), which transits to in-plane easy axis when CoFeB thickness increases as shown in Fig. 3(b). These samples were annealed 10min at 300°C. The interfacial anisotropy energy density ($K_i$) is determined to be 0.84 erg/cm$^3$ as shown in Fig. 3(c). However, the PMA of the Ag/CoFeB/MgO structure drop to nearly zero after the 400°C annealing. These results, together with other $K_i$ values obtained by our group with different HMs, are plotted in Fig. 3(d). The samples were typically annealed under two conditions. The "mild annealing" denotes the annealings carried out at 300°C for 10min to 60min. The "strong annealing" demotes annealings carried out at 400°C for 60min or longer. Note as a thermally activated process, the most pronounced change in MTJ properties occurs at the beginning of the annealing at high temperature as revealed previously.[26] The theoretically predicted $K_i$ values are also plotted in Fig. 3(d).[45] The calculated formation energy adapted from Ref. 43 are plotted in Fig. 3(e). A few important observations can be made from these data. The $K_i$ values after the mild annealing tend to agree reasonably well with the DFT calculation as seen in Fig. 3(d). However, the Ki values for some HM such as Hf and Ir fall dramatically to nearly zero after strong annealing. This behavior highlights the fact that though DFT calculations are very useful in predicting $K_i$, they are not applicable when the structure integrity of the HM/CoFeB/MgO stack no longer holds after strong annealings. Instead, the Ki values after strong annealing have a good correlation with the Fe-HM formation energy as shown in Fig. 3(e). Here we only focus on Fe since the composition of the FM layer is $Co_{20}Fe_{60}B_{20}$. The large drop of $K_i$ for Hf in particular is well explained for the very low Hf-Fe formation energy. The outlier in this group of HMs is Ag, where PMA is lost after strong annealing but the formation of energy of Ag-Fe is very high. A review on the binary phase diagrams shows that though all these HMs are immiscible with Fe as expected, Ag-Fe phase diagram has no intermediate phase as opposed to other HMs in this group. Therefore, results here indicate that the formation of intermediate compound between a HM and Fe is beneficial to reduce the nanoscale intermixing of the HM and Fe during annealing.

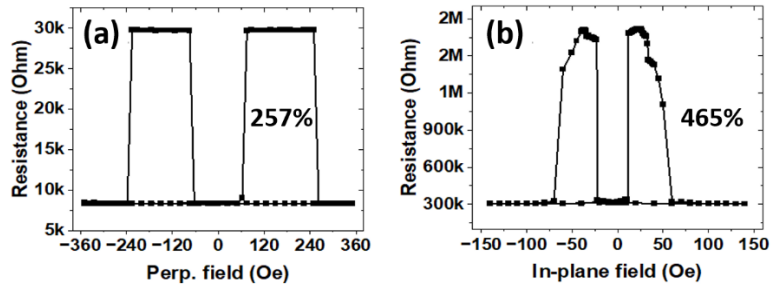

Figure 4. TMR curves of optimized MTJs with perpendicular magnetic anisotropy and in-plane magnetic anisotropy, respectively.

Finally, all the information obtained in this work are put together to optimize the TMR. The pMTJ stacks are with $HM_{BOT}$ and $HM_{TOP}$ as Ta(3.5nm)/Mo (0.8nm) and Mo (1.9nm)/Ta (4.5nm), respectively. The MgO was deposited at 255V as determined in Fig. 1(b). The utilization of a thin Mo layer enhances thermal stability, while allows Boron effectively passing through to be absorbed by the Ta layer.[20] This combination of Ta/Mo is able to produce high PMA as shown in Fig. 3(d), as well as a good (100) texture for the CoFeB layer as a thicker Mo layer tends to show (110)-oriented texture that may impede the solid state epitaxy of CoFeB(100) layers. The strong thermal robustness is evident from the good AP state of the pMTJ after the annealing at 500°C for 10min. As a results, a TMR of 257% is obtained in this device, which to the best of our knowledge is the highest value so far in voltage controllable MTJs. At the same time, a TMR of 465% was achieved in an in-plane MTJ with the successfully reduction of interfacial FeO layer. The bottom and top CoFeB thickness for the in-plane junction is 6nm and 4nm, respectively. Note the tuning of interfacial oxidation level described here only applies to the RF sputtering of MgO. If the MgO barrier is fabricated by other methods ( such as nature oxidation or reactive DC sputtering), different approaches are needed.

To conclude, the symmetry-conserved tunneling in MgO-pMTJs are strongly influenced by the interfacial oxidation level and thermodynamics of the HM/CoFeB/MgO structures. Though insertion of a Mg dusting layer was found to be not helpful, tuning the MgO adatom energy can effectively reduce the formation of the FeO layer thus increase the TMR. Through the comparison of different HM/CoFeB/MgO stacks, a high HM-Fe enthalpy of formation and the existence of HM-Fe intermediate phases are believed to be beneficial to the PMA and TMR. These results provide new means to achieve smaller, faster pMTJs that can be switched with lower energy.

**Acknowledgments**

This work was supported in part by Semiconductor Research Corporation through the Logic and Memory Devices program, by DARPA through the ERI program (FRANC), and by NSF through DMR-1905783. M. A and A.E. were supported by the REU supplement of NSF ECCS-1554011.

**Data availability**

The data that support the findings of this study are available from the corresponding author upon reasonable request.